\documentclass{article}

\usepackage[english]{babel}

\usepackage[letterpaper,top=2cm,bottom=2cm,left=3cm,right=3cm,marginparwidth=1.75cm]{geometry}

\usepackage{amsmath}
\usepackage{amssymb}
\usepackage{graphicx}
\usepackage{natbib}
\usepackage[colorlinks=true, allcolors=blue]{hyperref}

\newcommand\icarus{Icarus}% % Icarus
\newcommand\mnras{MNRAS}%   % Monthly Notices of the RAS
\newcommand\aj{{AJ}}%        % Astronomical Journal 
\newcommand\araa{{ARA\&A}}%  % Annual Review of Astron and Astrophys 
\newcommand\apj{ApJ}%    % Astrophysical Journal ++
\newcommand\apjl{{ApJL}}     % Astrophysical Journal, Letters
\newcommand\nat{{Nature}}%  % Nature
\newcommand\planss{{Planet.~Space~Sci.}}%  % Planetary Space Science
\newcommand\aap{{A\&A}}%     % Astronomy and Astrophysics 

\title{New results on orbital resonances}
\author{Renu Malhotra\\
Lunar and Planetary Laboratory, The University of Arizona \\ Tucson, Arizona 85721, USA \\ email: {\tt malhotra@arizona.edu} }

\begin{document}
\maketitle

\begin{abstract}
Perturbative analyses of planetary resonances commonly predict singularities and/or divergences of resonance widths at very low and very high eccentricities. We have recently re-examined the nature of these divergences using non-perturbative numerical analyses, making use of Poincar\'e sections but from a different perspective relative to previous implementations of this method. This perspective reveals fine structure of resonances which otherwise remains hidden in conventional approaches, including analytical, semi-analytical and numerical-averaging approaches based on the critical resonant angle. At low eccentricity, first order resonances do not have diverging widths but have two asymmetric branches leading away from the nominal resonance location. A sequence of structures called ``low-eccentricity resonant bridges" connecting neighboring resonances is revealed. At planet-grazing eccentricity, the true resonance width is non-divergent. At higher eccentricities, the new results reveal hitherto unknown resonant structures and show that these parameter regions have a loss of some -- though not necessarily entire -- resonance libration zones to chaos. The chaos at high eccentricities was previously attributed to the overlap of neighboring resonances. The new results reveal the additional role of bifurcations and co-existence of phase-shifted resonance zones at higher eccentricities. By employing a geometric point of view, we relate the high eccentricity phase space structures and their transitions to the shapes of resonant orbits in the rotating frame. We outline some directions for future research to advance understanding of the dynamics of mean motion resonances.
\end{abstract}

\section{Introduction}\label{s:intro}

Orbital resonances have influenced the properties and distribution of planets and minor planets in the solar system as well as in exo-planetary systems. Their study has a long history in celestial mechanics, physics and mathematics. There are many examples of stable mean motion resonances (MMRs) in our solar system, such as the Hilda group and the Trojan group of asteroids in 3/2 and 1/1 MMRs with Jupiter, the 2/3 MMR of Pluto with Neptune, the 5/2 near-resonance between Jupiter and Saturn, and the chain of 2/1 MMRs amongst the Galilean moons Io, Europa and Ganymede. 
Unstable mean motion resonances are also of great significance; for example, the Kirkwood Gaps in the asteroid belt are linked to the chaotic MMRs of Jupiter \citep[e.g.][]{Moons:1996}, and the long term stability of the solar system may be linked to the role of MMRs \citep[e.g.][]{Murray:1999,Guzzo:2006}. There is also a growing body of literature on the importance of MMRs in the dynamics of exoplanetary systems \citep[recent review by][]{Zhu:2021}. %\citep[e.g.][]{Marcy:2001,Lee:2002,Thommes:2008,Lithwick:2012,Petrovich:2013,Fabrycky:2014,Goldreich:2014,Mills:2016,Zhu:2021}.

Previous analytical as well as numerical treatments of MMRs have been based on or informed by perturbation theory. Commonly, these treatments have invoked the averaging principle to discard the short timescale terms to reduce the problem to one degree of freedom by identifying a slow timescale and a corresponding ``critical resonant angle", and developing a pendulum-like model for the dynamics in that single degree of freedom. The common pendulum is the first approximation for describing the dynamics of MMRs. In this model, the phase space is divided into a zone of oscillations of the critical resonant angle (at sufficiently low relative energies), and rotation zones (both clockwise and counter-clockwise rotations, at sufficiently high energies); the boundaries between the oscillations and the rotations is a separatrix of zero frequency. For elementary expositions, the common pendulum model suffices \citep{Dermott:1983,Winter:1997}. The ``second fundamental model of resonance" (SFMR) introduced by \cite{Henrard:1983} provides a more accurate description and it has been a mainstay of such analytical treatments of MMRs. Similar to the common pendulum, the SFMR is also of one degree of freedom, but, unlike the common pendulum, its potential function is not independent of the canonical momentum. Specifically in the case of planetary mean motion resonances, the perturbative potential function possesses the d'Alembert characteristic in $(\sqrt{J},\phi)$, where the canonical coordinate, $\phi$, is identified with the critical resonant angle and $J$ is its canonically conjugate momentum (obtained by means of canonical transformations starting from the Delaunay elements for the unperturbed Kepler problem). %These analytical treatments stimulated several insightful analytical and numerical studies of specific cases of mean motion resonances in the context of the solar system. \citet{Wisdom:1980,Henrard:1983b,Wisdom:1985,Moons:1996} and \citet{Winter:1997} investigated the formation and extent of the Kirkwood Gaps in the main asteroid belt. \citet{Morbidelli:1995} and \citet{Malhotra:1996} investigated the phase space structure near Neptune's mean motion resonances in the Kuiper belt. 

While the above-mentioned approaches mitigated the problem of ``small divisors" in historical perturbation theory treatment of mean motion resonances, they still retained certain features and ambiguities that were puzzling to this author; in particular, the widths of mean motion resonances exhibited singularities and/or divergences at low eccentricities. For example, the textbook by \cite{Murray:1999SSD} gives an analytical estimate that the resonance zone widths of Jupiter's first order MMRs in the asteroid belt diverge to infinity on both sides of the nominal resonance location as an asteroid's eccentricity approaches zero (their Eq.~8.76 and Fig.~8.7). \cite{Wisdom:1980} used an analytical perturbative approach and \cite{Winter:1997} used a non-perturbative numerical approach with Poincar\'e sections to measure the widths of Jupiter's resonances in the asteroid belt, and they also reported a similar divergence, albeit on only one side of the nominal resonance location. On the same topic, the textbook by \cite{Morbidelli:2002Book} also describes that the stable resonance center of Jupiter's 2/1 MMR diverges only on one side of the nominal resonance location as $e$ approaches zero, and that a resonance separatrix ``vanishes" for $e\lesssim0.2$ making the resonance width undefined for smaller eccentricities (his Fig.~9.11).

At higher eccentricities, previous studies either extrapolated the low-eccentricity analytical models or used a numerical approach to average the perturbation potential over the fast degrees of freedom. Such studies reported the divergence of resonance widths near planet-grazing eccentricities and a gradual decrease of resonance widths at higher eccentricities \citep{Moons:1993,Morbidelli:1995,Nesvorny:1997,Deck:2013,Hadden:2018b,Gallardo:2019}.

The high eccentricity regime of planetary mean motion resonances has gained increasing attention as discoveries of minor planets and dwarf planets in the distant solar system show that this regime of phase space does have a significant influence on the dynamics of these objects. For example, the phenomenon of long term ``resonance sticking" is commonly observed, typically at high eccentricities, in numerical studies of a class of Kuiper belt objects called the ``scattered disk" or the ``scattering disk" \citep{Duncan:1997,Lykawka:2007a,Gladman:2008,Gladman:2012,Yu:2018}. Shorter term resonance sticking is reported amongst the Centaurs \citep{Belbruno:1997,Tiscareno:2003,Bailey:2009,Fernandez:2018,Roberts:2021}. Recent interest in the possible observable effects of an unseen distant planet in the solar system have also stimulated interest in the dynamics of high eccentricity MMRs~\citep[e.g.,][]{Malhotra:2016a,Beust:2016,Hadden:2018a}. %Many numerical simulations have probed the high eccentricity regime of MMRs, but there have been comparatively few attempts at theoretical understanding. 

The desire to understand better the high eccentricity regime of MMRs as well as the puzzling features at low and moderate eccentricities was the motivation for a sequence of recent investigations in my research group \citep{Wang:2017,Malhotra:2018a,Lan:2019,Malhotra:2020}. In these studies, we adopted a non-perturbative approach by computing Poincar\'e sections, but with some modifications relative to previous studies. The modifications have helped to significantly clarify some aspects of the dynamics of MMRs. Two additional recent publications, \cite{Lei:2020} and \cite{Antoniadou:2021}, have elaborated further on some of these results.
The present article summarizes a part of these advances, specifically those related to interior mean motion resonances of the first order, that is, when the orbital periods of two planets orbiting a central host star are close to the ratio of two small integers differing only by unity, e.g., 2/1, 3/2, etc., and the interior planet is of negligible mass. 

Results of our investigations of exterior MMRs can be found in \cite{Malhotra:2018a}, \cite{Lan:2019}, and are summarized in \cite{Malhotra:2019b}. We also refer the reader to \cite{Deck:2013}, \cite{Ramos:2015}, \cite{Hadden:2018b}, \cite{Hadden:2019}, and \cite{Petit:2021} for additional new results on mean motion resonances, including in the non-restricted, non-circular co-planar three body model and on three-planet mean motion resonances; these are beyond the scope of the present review.

\section{A new approach to Poincar\'e sections}\label{s:new-Poincare}

We begin with the physical model known as the restricted three body problem, and its mathematically simplest version which has the massive planet (``the perturber"), of mass $m_2=\mu$, in a circular orbit of unit radius about the host star whose mass is $m_1=1-\mu$. This model admits an integral of the particle's motion, the Jacobi integral $C_J = 2(\Omega L-E)$, where $\Omega$ is the angular velocity of the massive bodies in their circular orbits and $E$ and $L$ are the specific energy and specific angular momentum of the particle (in the barycentric frame, with $z$-axis oriented along the total angular momentum of the massive bodies).  A further simplification is that the zero-mass particle is restricted to move in the planet's orbital plane; this is the planar circular restricted three body problem (PCRTBP). Although it does not describe the full complexities of real planetary systems, this model remains a very useful approximation and provides insights into the phase space structure of MMRs. The model has two degrees of freedom, hence a phase space of four dimensions. The existence of the Jacobi integral implies that the motion of the test particle takes place on a three dimensional surface in the four dimensional phase space. This permits the visualization of the phase space structure in two dimensional surfaces called ``surfaces of section" or Poincar\'e sections, devised by the mathematician Henri Poincar\'e (1854-1912). These are akin to stroboscopic plots with which we can track the motion of the particle in two variables to visualize on a flat plane. 

\begin{figure}[ht]
 \centering
\hspace{25pt} \includegraphics[width=100mm] {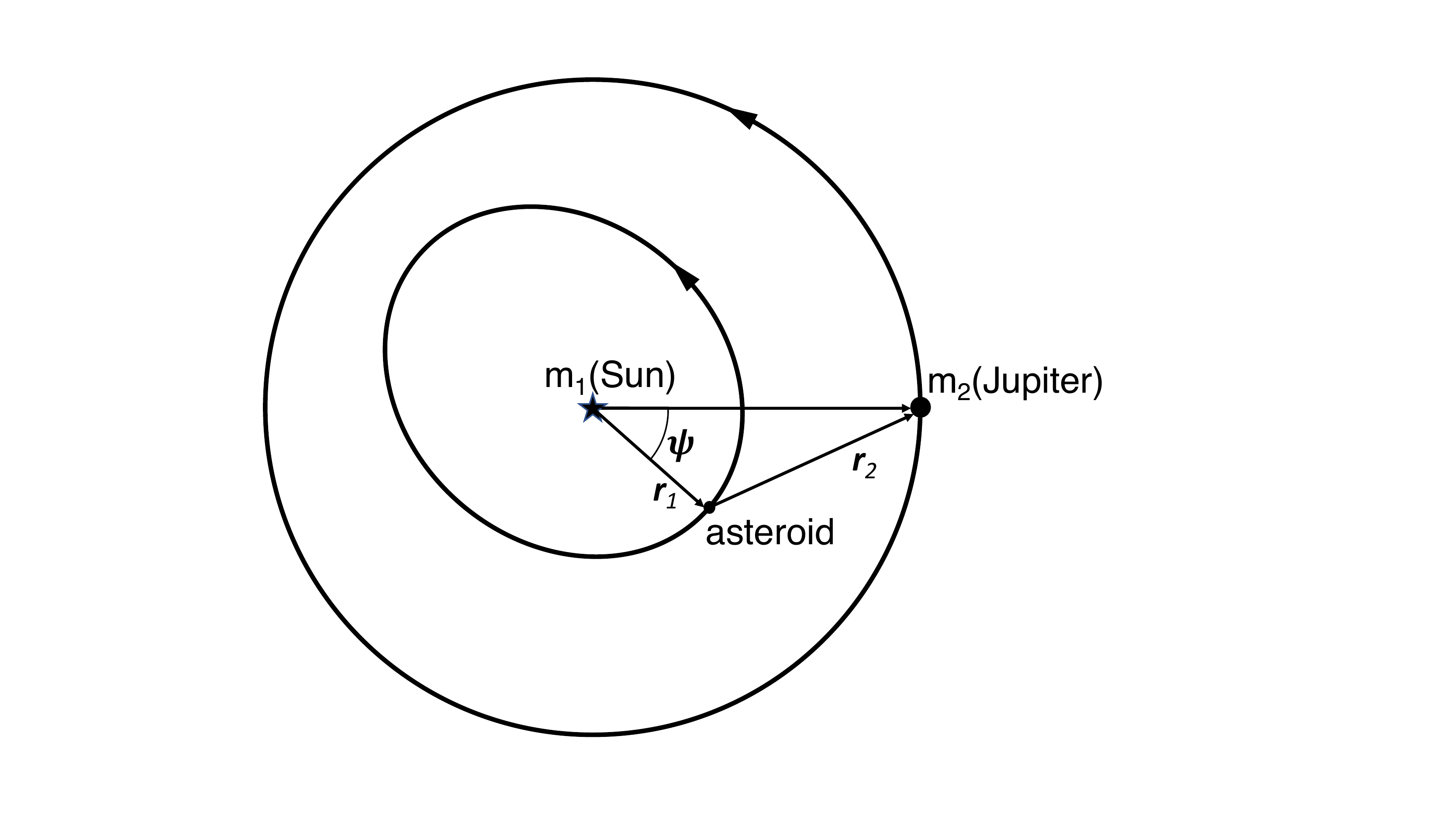} %{/Users/renu/Projects/NanZhang/Paper_mnras/arxiv/f1.pdf}
 \caption{A schematic diagram to illustrate the definition of $\psi$, the angular separation of Jupiter from the test particle when the latter is at pericenter.}
 \label{f:f1}
\end{figure}

Previous studies commonly followed the lead of \cite{Henon:1966} on the choice of Poincar\'e sections for the PCRTBP by recording the test particle's state vector at every successive conjunction (or opposition) with the perturber \citep[e.g.][]{Duncan:1989,Winter:1997}). With this choice, the Poincar\'e sections were presented as plots of $\dot x$ vs.~$x$, where $(x,y)$ and $(\dot x,\dot y)$ are the cartesian position and velocity components in the synodic frame, that is, the barycenteric frame co-rotating with $m_1$ and $m_2$; the primaries' locations in this frame are fixed on the $x$-axis at $-\mu$ and $1-\mu$, respectively. However this is not a unique choice. 
In our implementation, we made two modifications relative to previous studies:

\begin{itemize}
\item  First, we recorded the test particle's state vector at every successive perihelion passage. This choice is physically motivated so as to trace the behavior (libratory or not) of the test particle's perihelion longitude relative to the perturbing planet's position. 
\item  Second, we recognized that there were other possible choices of dynamical variables for the Poincar\'e plots than the rotating frame configuration variables, $(x,\dot x)$, most commonly adopted in previous studies. With the stroboscopic record of the state vector at successive perihelion passages, one could examine many different combinations of dynamical variables. Most useful are the two-dimensional plots in $(x,y)$, $(e\cos\psi,e\sin\psi)$, and $(\psi,a)$, where $x,y$ are the cartesian position coordinates in the rotating frame, $a$ and $e$ are the osculating semimajor axis and eccentricity of the particle, and $\psi$ measures the angular separation of the perturber from the test particle when the particle is at perihelion. The definition of $\psi$ is illustrated in Figure~\ref{f:f1} in which the perturber is Jupiter and the test particle is an asteroid.
\end{itemize}

%%%%%%%%%%%%%%%%%%%%%%%%%%%%%%%%%%%%%%%%%%%%%%%%%%%%
% 
%From Malhotra:2020
%%%%%%%%%%%%%%%%%%%%%%%%%%%%%%%%%%%%%%%%%%%%%%%%%%%%

These modifications enable a visualization of the phase space structure from a different point of view than in previous studies. In the previous implementations of Poincar\'e sections of the PCRTBP,  only one point is recorded per synodic period, whereas in our implementation, typically there are multiple points recorded per synodic period. For example, in the neighborhood of an interior $(p+1)/p$ resonance, there are $(p+1)$ points recorded per synodic period. In this sense, our implementation gives a higher resolution view of the phase space structure than in the previous approach. The new perspective has helped to resolve the puzzles and ambiguities of resonance widths mentioned above (Section~\ref{s:intro}).

In addition, we examine the shape of the resonant orbit in the rotating frame and how it changes with increasing (or decreasing) value of the particle's eccentricity. And we correlate its shape with the evolving topology of the phase space as visible in the Poincar\'e sections (Section~\ref{s:geometry}). The exercise of correlating the phase space structure with the geometry of the resonant orbit yields insights into the physical origins of the multiple resonance libration modes in the moderate-to-high eccentricity regime: it explains both their existence and their size as a function of the particle's eccentricity. It also reveals the physical connections between resonance zones and stable/unstable periodic orbits of the PCRTBP. Periodic orbits present as fixed points in the Poincar\'e sections and define the centers and boundaries of the resonance zones.

The relationship between the conventional approach taken in analytical and numerical-averaging approaches and this different perspective is actually rather simple.  For a first order MMR, the usual ``critical resonant angle" is defined as
\begin{equation}
\phi = (p+1)\lambda' - p\lambda - \varpi,
\label{e:phi}\end{equation}
where $p$ is an integer, $\lambda,\lambda'$ are the mean longitudes of the particle and of the perturbing planet, respectively, and $\varpi$  is the particle's longitude of perihelion. In our definition of the Poincar\'e section, the particle is located at its perihelion, that is, when $\lambda=\varpi$. It then follows that $\phi$ and $\psi$ are related:
\begin{equation}
\phi = (p+1)\psi .
\label{e:phi-psi}\end{equation}
It must be understood that $\phi$ is a continuous function of time, whereas $\psi$ is a stroboscopic coordinate, being defined at the point in time when the particle is at perihelion. The choice of $\psi$, together with our non-perturbative approach, reveals more details of the phase space structure than have been possible with the conventional use of $\phi$ in analytical, semi-analytical and numerical-averaging approaches in the previously existing literature. In particular, we will see below that the use of $\psi$ proves to be the key to solving one of the puzzles in the previous literature, that of the apparent ``vanishing of the separatrix'' at low eccentricity.

%%%%%%%%%%%%%%%%%%%%%%%%%%%%%%%%%%%%%%%%%%%%%%%%%%%%
%%%%%%%%%%%%%%%%%%%%%%%%%%%%%%%%%%%%%%%%%%%%%%%%%%%%
% From Malhotra:2020
% f2ca.png = 3.154a.png  f2cb.png = 3.154e.png   f2cc.png = 3.154xy.png
\begin{figure*}[b]
\centering
   \begin{tabular}{c c c}
     \includegraphics[width=1.57in]{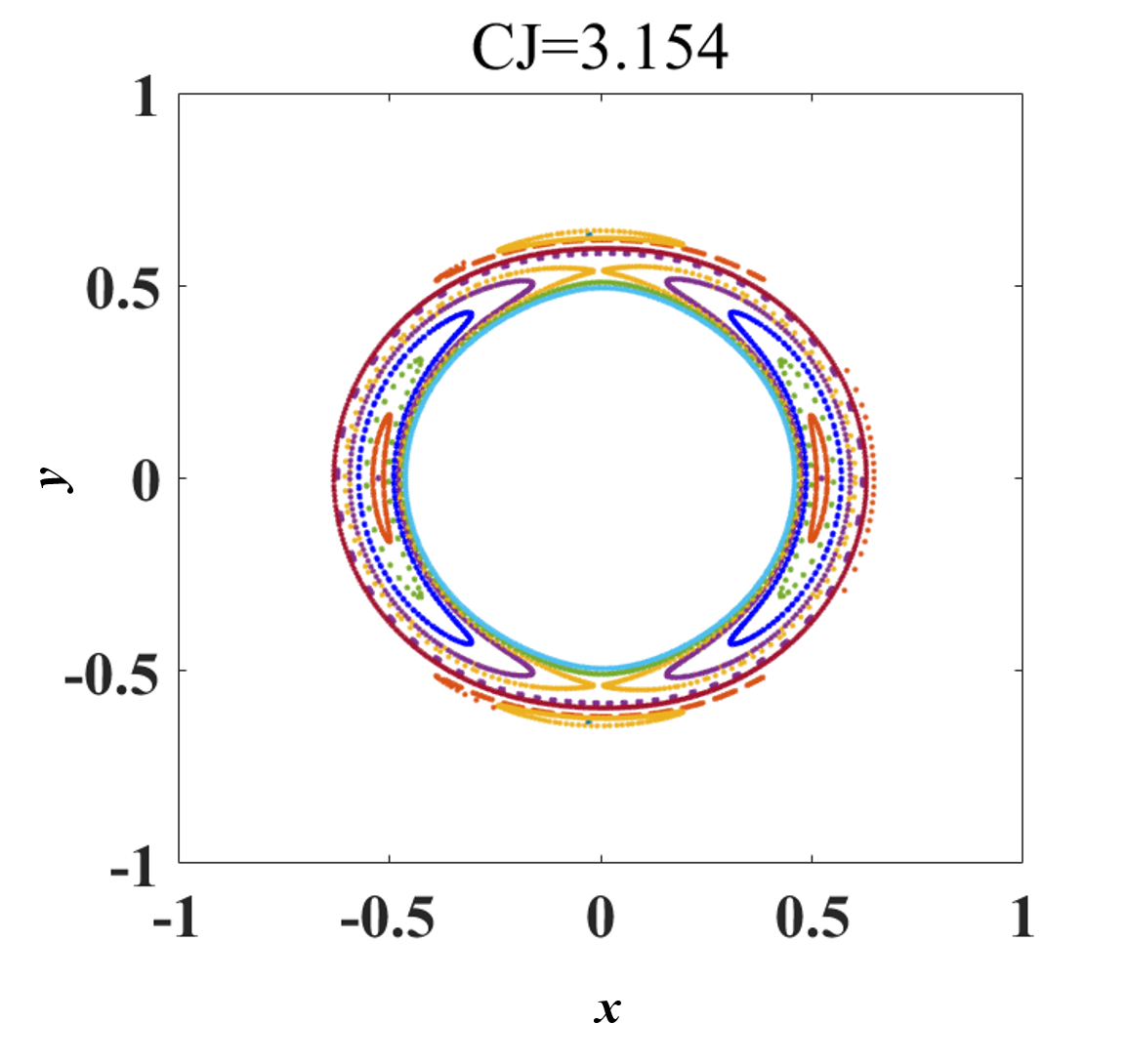} %{/Users/renu/Projects/NanZhang/Paper_mnras/arxiv/f2cc.png}&
      \includegraphics[width=1.97in]{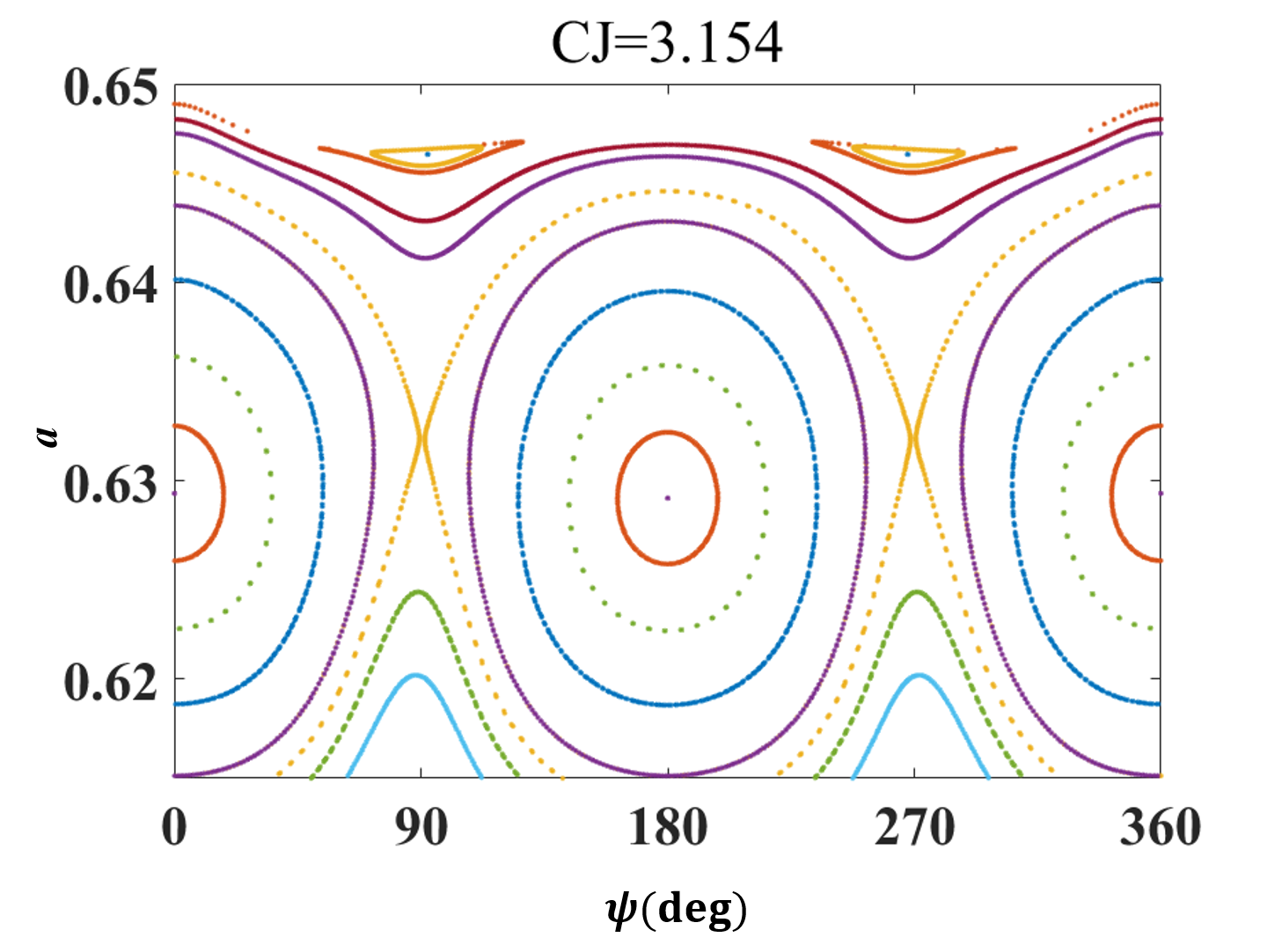} %{/Users/renu/Projects/NanZhang/Paper_mnras/arxiv/f2ca.png} &
   \includegraphics[width=1.57in]{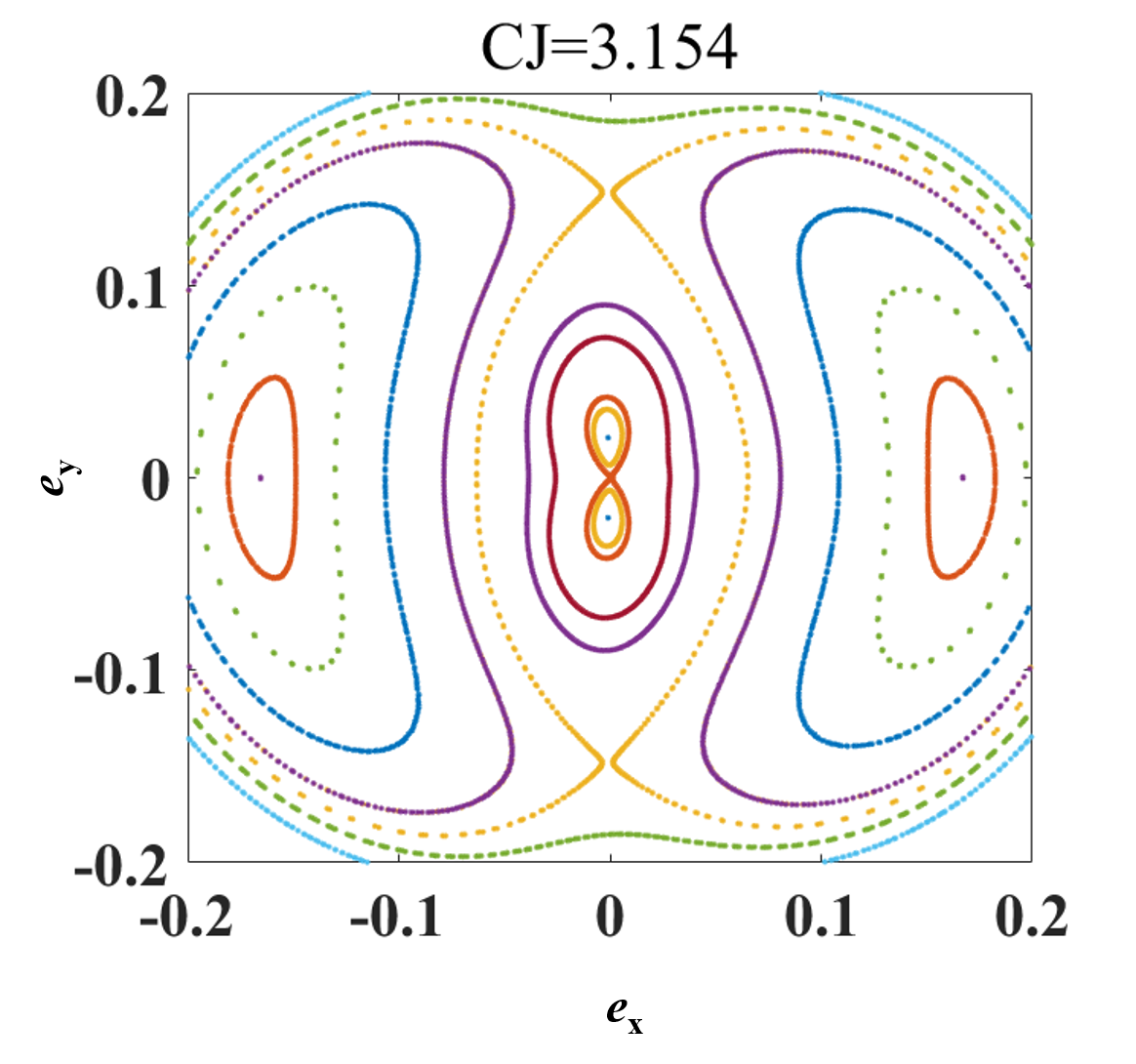} %{/Users/renu/Projects/NanZhang/Paper_mnras/arxiv/f2cb.png} \\
   \end{tabular}   
 \caption{Poincar\'e sections near Jupiter's 2/1 interior resonance for one value of the Jacobi constant, as indicated by the legend at the top of each panel. 
 The left panel displays the Poincar\'e section in the configuration space, $(x,y)$, in the rotating frame; the locations of the Sun and Jupiter are fixed in this plane at $(x,y)=(-\mu,0)$ and $(x,y)=(1-\mu,0)$, respectively. The middle panel plots the same Poincar\'e section in the $(\psi,a)$ plane, and the right panel in the $(e_x,e_y)\equiv(e\cos\psi,e\sin\psi)$ plane.}
 \label{f:f2}
\end{figure*}
%%%%%%%%%%%%%%%%%%%%%%%%%%%%%%%%%%%%%%%%%%%%%%%%%%%%

Figure~\ref{f:f2} shows an example set of Poincar\'e sections obtained with the above approach for one value of the Jacobi integral in the neighborhood of Jupiter's 2/1 interior MMR. In the $(x,y)$ plane, we observe that there are two pairs of libration zones, one pair is centered on the $x$-axis and the other pair is centered (almost) on the $y$-axis. For the former, the test particle's alternate perihelion passages occur at conjunction and opposition with Jupiter; these can be called the pericentric resonant islands. For the latter, the alternate perihelion passages occur close to $90^\circ$ leading and trailing Jupiter's longitudinal position; consequently, conjunctions and oppositions with Jupiter occur at the particle's aphelion, so these can be called the apocentric resonant islands. In the $(\psi,a)$ plane as well as in the $(e\cos\psi,e\sin\psi$) plane, the same two pairs of libration islands are visible, with the first pair (centered at $\psi=0$ and $\psi=180^\circ$) being the larger size and at larger eccentricity, while the second pair (centered near $\psi=90^\circ$ and $\psi=270^\circ$) being the smaller size and at smaller eccentricity. It is striking that the larger pair of islands has a central value of $a$ close to (slightly below) the nominal resonant value of $a=(1/2)^{\frac{2}{3}}=0.630$, whereas the smaller pair is displaced to a higher value.  (The keen reader will also notice that the centers of the smaller islands are also visibly displaced slightly from $\psi=90^\circ,270^\circ$; this is discussed further below.) The existence of two pairs of resonant islands illustrates the existence of two branches of the resonance: the pericentric branch in which the conjunctions of the particle and Jupiter librate about the particle's pericenter, and the apocentric branch in which the conjunctions librate about the particle's apocenter. 

We point out that in the example shown in Figure~\ref{f:f2}, there are two different separatrices visible in the Poincar\'e sections. One separatrix delineates the boundaries of the pericentric resonant islands, the other the boundaries of the apocentric resonant islands. The separatrix delineating the apocentric librations (at the smaller eccentricity) has not been unambiguously identifiable in the perturbative treatments in previous studies. For example, in the second fundamental model of resonance, SFMR, there is only one separatrix, and we can identify it with the separatrix of the pericentric librations in Fig.~\ref{f:f2}; although the apocentric libration zone exists in the SFMR, its boundary is not clearly a zero-frequency separatrix in that model. This problem carries over into semi-analytical models employing numerical averaging which, in common with the SFMR, also employ the critical resonant angle, $\phi$, as the single degree of freedom. This is the reason for the apparent ``vanishing of the separatrix'' at low eccentricity claimed in previous studies, such as in \citet{Morbidelli:2002Book} and in \citet{Ramos:2015}.  %By relying on the critical resonant angle, $\phi$, the SFMR does not distinguish amongst the multiple resonant islands that exist in the physical librations of $\psi=\phi/(p+1)$. The identification of $\psi$ as the dynamical variable to more accurately describe resonance dynamics is a simple yet important advance on this topic.

The exclusive use of the critical resonant angle in previous studies also suppresses the distinction amongst the different islands of the pericentric branches of the resonance. As illustration, we observe in Fig.~\ref{f:f2} that the two pericentric islands in the Poincar\'e section for Jupiter's 2/1 interior MMR are not exactly symmetric: the value of $a$ at the center of the island at $\psi=0$ is slightly larger than that of the island at $\psi=180^\circ$. The asymmetry is also evident in the  shapes of the boundaries of the two islands in the $(\psi,a)$ plane. Consequently, when we measure the extent of the resonance zone in semi major axis (discussed in Section \ref{s:res-width}), we find that each of the two islands has slightly different values of $a_{\mathrm{min}}$ and $a_{\mathrm{max}}$, resulting in two different estimates of the resonance width. 

In general, for interior first order $(p+1)/p$ MMRs, these Poincar\'e sections show a chain of $(p+1)$ islands for the pericentric branch and a chain of $(p+1)$ islands for the apocentric branch. Likewise, for exterior first order $p/(p+1)$ MMRs, these Poincar\'e sections will show a chain of $p$ islands for the pericentric branch and a chain of $p$ islands for the apocentric branch. For example, Jupiter's interior 3/2 MMR in the asteroid belt appears as a chain of three islands for the pericentric branch and another chain of three islands for the apocentric branch; Neptune's exterior 2/3 MMR in the Kuiper belt appears as a chain of two islands for the pericentric branch and another chain of two islands for the apocentric branch.  

There are some ranges of the Jacobi integral for which one of the branches vanishes. For example, in the low eccentricity regime for interior MMRs, it is the apocentric branch that does the vanishing, and for exterior MMRs it is the pericentric branch that does the vanishing trick. These branches reappear at higher eccentricities, and, in some cases, vanish and reappear multiple times at high eccentricities. This is discussed further in the following sections.

\section{Resonance Widths}\label{s:res-width}

\subsection{Measuring resonance width}\label{ss:measure-res-width}

%%%%%%%%%%%%%%%%%%%%%%%%%%%%%%%%%%%%%%%%%%%%%%%%%%%%
% Malhotra:2020
\begin{figure*}[b!]
 \centering
 \vglue-1.2in \includegraphics[angle=270,width=5.0in]{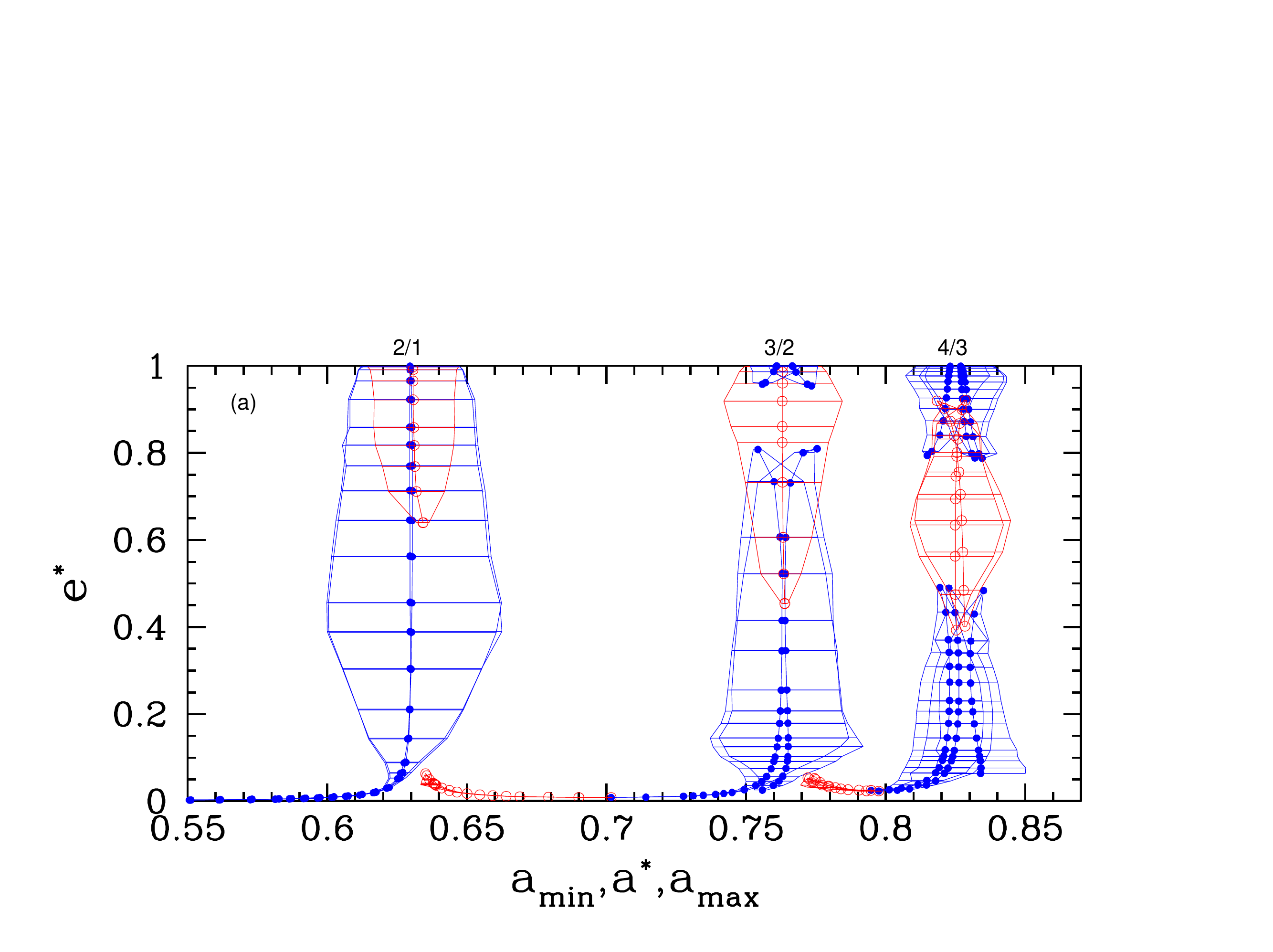} 
  \vglue-1.3in  \includegraphics[angle=270,width=5.0in]{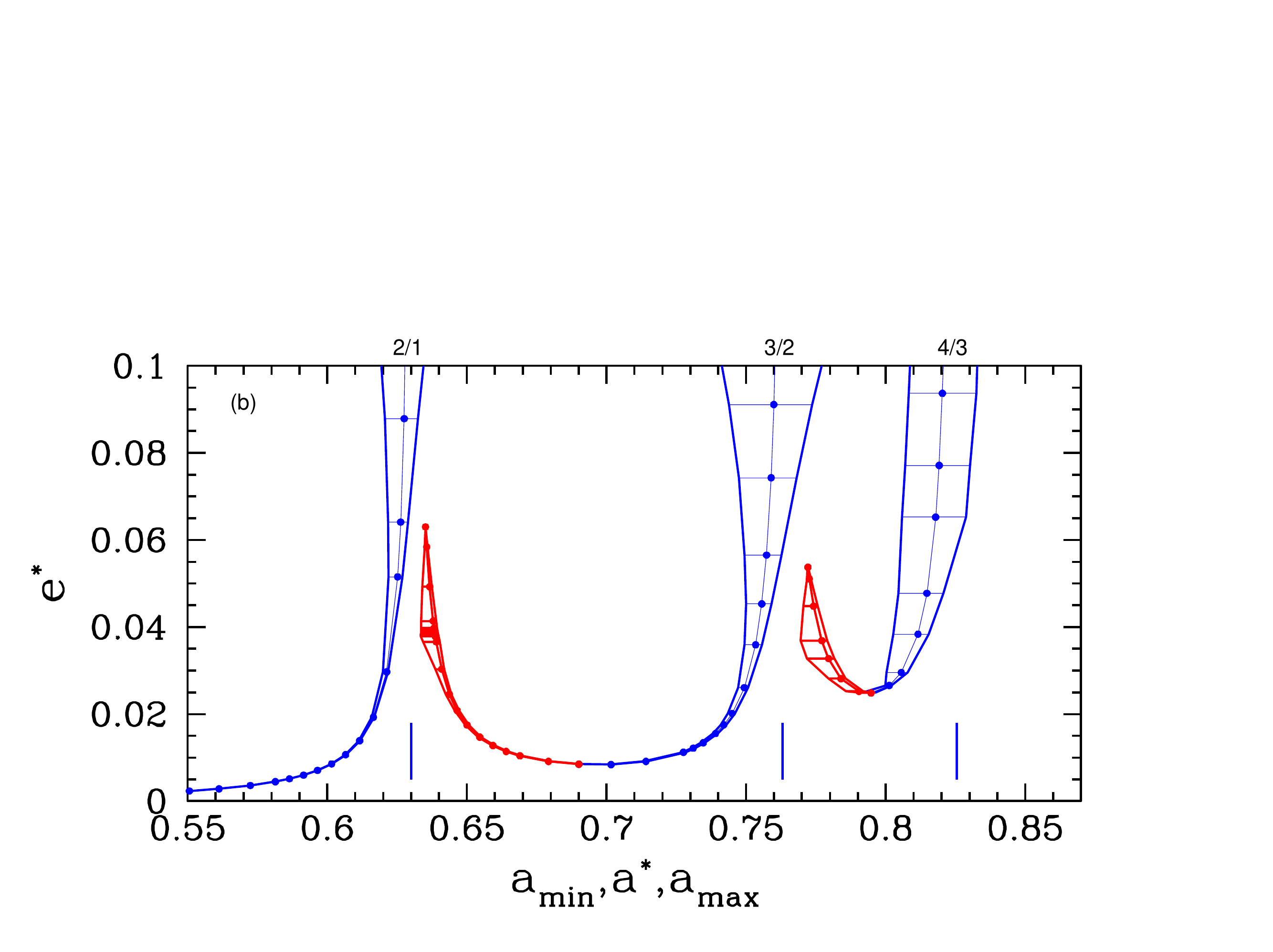}
 \caption{(a) The libration centers and widths of Jupiter's 2/1, 3/2 and 4/3 interior resonances for the full range of eccentricities, $0<e<1$, for the pericentric and apocentric libration islands, in blue and red, respectively. The center $(a^*,e^*)$ of the pericentric libration islands is indicated by the filled circles, that of the apocentric libration islands is indicated by the open circles. The maximum libration range of $a$ in each case is indicated by the horizontal bars. 
 (b) Expanded view of one of each of the pericentric and apocentric branches at low eccentricity. The pericentric island shown is centered near $360^\circ/(p+1)$ while the apocentric island is centered near $180^\circ/(p+1)$ where $p=1,2,3$ for the 2/1, 3/2, 4/3 MMRs, respectively. %For the 2/1 MMR, the representative pericentric island shown is centred at $\psi=180^\circ$ and the representative apocentric island shown is centred at $\psi=90^\circ$. For the 3/2, the pericentric island shown is centred near $\psi=120^\circ$ and the apocentric island is centred near $\psi=60^\circ$. For the 4/3, the pericentric island shown is centred near $\psi=90^\circ$; 
 The apocentric branch of the 4/3 is not shown because it has dissolved into a chaotic zone. The short vertical lines indicate the nominal location, $a_{\mathrm{res}} = (p/(p+1))^{\frac{2}{3}}$.
}
 \label{f:f3}
\end{figure*}
%%%%%%%%%%%%%%%%%%%%%%%%%%%%%%%%%%%%%%%%%%%%%%%%%%%%

To measure the resonance width in semi major axis as a function of eccentricity, we first identify the center of the resonance by the values $a^*$ and $e^*$ at the center of each resonant island. For $a^*$, we read this central value in the $(\psi,a)$ plane, and for $e^*$, we read it in the $(e\cos\psi,e\sin\psi)$ plane. Then we measure the minimum and maximum values of $a$ in the libration zone of each resonant island in the $(\psi,a)$ plane. For some ranges of the Jacobi integral, the resonant islands may not possess a regular separatrix but present a fuzzy boundary due to a chaotic zone. In these cases we estimate values of $a_{\mathrm{min}}$ and $a_{\mathrm{max}}$ for the largest visible regular librating trajectory in the $(\psi,a)$ plane. This procedure is rather labor-intensive, and the resolution in $a,e$ that has been achieved in the results reported thus far is certainly limited by the available time and capacity of the researchers.

Using this procedure, we recently investigated several first order interior MMRs of Jupiter (as well as lower-mass perturbers) and many exterior MMRs of Neptune \citep{Wang:2017,Lan:2019,Malhotra:2020}. The correlated measure, $\Delta e$, of the range of eccentricity libration accompanying the maximal range, $\Delta a$, is additionally reported in \cite{Lei:2020}. Here we illustrate the main new results with a few examples.
The numerically determined resonance centers and widths of the 2/1, 3/2 and 4/3 interior MMRs of Jupiter are shown in the $(a,e)$ parameter space in Figure~\ref{f:f3}(a) for the entire eccentricity range, $0<e<1$; in Fig.~\ref{f:f3}(b), the low eccentricity regime, $0<e<0.1$, is shown in an expanded scale.  There are several interesting features in this figure, as we discuss below.

As mentioned above, a $(p+1)/p$ interior MMR presents in the Poincar\'e sections as a chain of $(p+1)$ islands each for the apocentric and pericentric branches. (In the terminology of period orbit theory, the centers of a chain of $p$ resonant islands represent a single stable periodic  orbit of period $p$.) 
Only those islands located symmetrically relative to the $x$-axis (equivalently, $\psi=0$) have identical central values, $a^*$ and $e^*$, and of the resonance width (measured with $a_{\mathrm{min}}$ and $a_{\mathrm{max}}$); the other islands have slightly different resonance centers and widths. This means that in the $(a,e)$ parameter space, the number of distinct resonance boundaries is not always equal to the number of resonant islands that appear in the Poincar\'e sections. For example, for Jupiter's 3/2 interior MMR, there are three resonant islands of the pericentric branch, centered at $\psi=0$ and near $\psi=120^\circ,240^\circ$, respectively. The latter two have the same values of $a^*$ and $e^*$ at their centers and the same width ($a_{\mathrm{min}}$ to $a_{\mathrm{max}}$) in semi major axis, but the one centered at $\psi=0$ differs slightly in both respects. (The magnitude of the difference depends upon the perturber's mass.) %For additional details, we refer the reader to \cite{Malhotra:2020}. 
In Fig.~\ref{f:f3}, for each of Jupiter's 2/1, 3/2 and 4/3 interior MMRs, we plot the center and widths of the pericentric islands in blue %(with center near $360^\circ/(p+1)$)
and the apocentric islands in red. %(with center near $180^\circ/(p+1)$). 
At the resolution of Fig.~\ref{f:f3}(a) the differences between the two pericentric islands of Jupiter's interior 2/1 MMR are not very visible, but the differences amongst the pericentric islands of the interior 3/2 and of the 4/3 are more visible. Such differences are also more evident amongst the multiple islands at higher eccentricities in Fig.~\ref{f:f3}(a).
%We note that the low eccentricity apocentric branches of Jupiter's interior 4/3 MMR are not shown in this figure; this is because the corresponding islands in the Poincar\'e sections of this parameter region are extremely small and invisible within a chaotic sea.

\subsection{Fine structure at low eccentricities}\label{ss:low-ecc}

For small values of eccentricity, the features of note in Fig.~\ref{f:f3}(b) are as follows: 
\begin{itemize}
\item[(i)] the pericentric and apocentric branches both exist continuously at significant distances from the nominal location of each MMR; 
\item[(ii)] the pericentric branches increase in width as $a^*$ approaches the nominal resonance location from the left (from lower values); and 
\item[(iii)] the apocentric branches first increase in width, then decrease and terminate as $a^*$ approaches the nominal resonance location from the right (from higher values).  
\end{itemize}

With these results we learn that the divergence of the resonance boundaries described in the previous literature \citep[e.g.][]{Wisdom:1980,Murray:1999SSD} is actually resolved into two branches of the resonance, each of which diverges away from the nominal resonance location, but the width of each branch does not diverge, rather it decreases as eccentricity decreases. Moreover,  the Poincar\'e sections near very small eccentricity show that the apocentric branch is bounded by a bonafide separatrix which passes through $e=0$. This separatrix is not visible in previous analytical treatments nor in the semi-analytical and numerical-averaging treatments because those approaches employed the critical resonant angle as the dynamical coordinate \citep{Henrard:1983b,Nesvorny:1997,Morbidelli:2002Book}. The choice of this coordinate hides the existence of the separatrix passing through $e=0$. This is the reason why the fine structure of first order resonances at low eccentricities has remained unresolved and unremarked in the previous literature on resonance widths. 

Our change of coordinate, from the critical resonant angle, $\phi$, to its sub-multiple, $\psi=\phi/(p+1)$ (cf.~Eq.~\ref{e:phi-psi}), is essential to reveal the fine structure at low eccentricity. This has also been demonstrated by \cite{Lei:2020} who recover most of the low-eccentricity fine structure of a first order MMR described in \cite{Malhotra:2020} within a semi-analytical approach parallel to the SFMR, but using $\sigma=\phi/(p+1)$ as the canonical coordinate. These authors further show that the SFMR, with its truncation of the resonant potential to only one leading-order term, is insufficient to describe accurately the dynamics near a first order MMR at low eccentricities (see also \cite{Beauge:1994}); at least two additional terms of the resonant perturbation potential are necessary to recover the topology of the phase space approximately consistent with the non-perturbative results of \cite{Malhotra:2020}.  Some features revealed with the non-perturbative approach nevertheless remain hidden in the semi-analytical results of \cite{Lei:2020}. This is due to the fundamental approximation in any analytical or semi-analytical approach that all but the specific resonant terms of the perturbation potential are discarded or numerically averaged out. One significant feature revealed uniquely in the non-perturbative approach is the existence of low eccentricity bridges between neighboring first order MMRs; this is described in the next section.

\subsection{Low eccentricity resonant bridges}\label{ss:bridges}

We point out one of the most interesting features of first order MMRs at low eccentricity: that the pericentric branch of the 4/3 MMR smoothly connects with the apocentric branch of the 3/2, and, similarly, the pericentric branch of the 3/2 smoothly connects with the apocentric branch of the 2/1. This can be seen in Fig.~\ref{f:f3}(a), and at higher resolution in Fig.~\ref{f:f3}(b). 
These connections between neighboring first order MMRs occur with a gradual transformation of the phase space structure as we tune the Jacobi integral across a range of values encompassing neighboring MMRs. For example, the transformation of the phase space structure from the chain of four resonant islands of the pericentric branch of the 4/3 MMR to the chain of three resonant islands of the apocentric branch of the 3/2 occurs as follows: as the Jacobi integral increases, the four pericentric islands decrease in size; two of these islands -- those centered near $90^\circ$ and $270^\circ$ -- gradually move their centers towards $60^\circ$ and $300^\circ$, respectively; one of the four islands, the one centered at $\psi=0$, migrates to smaller values of $e^*$, and its size decreases faster than that of the other three, until it vanishes. This leaves only three islands, of similar size, which then form the three apocentric islands of the 3/2 resonance. Similarly, the transformation of the chain of three resonant islands of the pericentric branch of the 3/2 MMR to the two-island chain of the apocentric branch of the 2/1 occurs by the gradual shrinking and disappearance of one of the three pericentric libration islands of the 3/2, the one centered at $\psi=0$, and the gradual migration of the other two islands -- those centered near $\psi=120^\circ$ and $\psi=240^\circ$ -- to positions near $\psi=90^\circ$ and $\psi=270^\circ$ of the 2/1 apocentric branch. These features have remained unremarked in the previous literature. 

We christened these features as ``resonant bridges" between first order MMRs, and conjectured that these structures could serve as long range transport conduits under weak dissipative forces, so that a particle could adiabatically move long radial distances along these resonant paths, including transferring amongst different MMRs.  
The study by \cite{Antoniadou:2021} confirms this conjecture, with the caveat that the dissipative forces be strong enough to avoid the traps of higher order MMRs in that path.  %These authors also point out that the libration centers of the low eccentricity pericentric and apocentric branches are part of the so-called circular families of periodic orbits of the PCRTBP.  We notice that the locations, $(a^*,e^*)$, of the pericentric and apocentric resonance branches in their analysis have some differences compared to those computed in our non-perturbative analysis with Poincar\'e sections, such as the $(a^*,e^*)$ locations of high-eccentricity apocentric branches. It would be interesting and useful to reconcile these differences.

Orbital migration and adiabatic evolution near individual MMRs has been a subject of many previous investigations. Applications include the phenomena of capture into resonance during convergent migration and eccentricity excitation during either convergent or divergent migration  \citep[e.g.,][]{Dermott:1988,Peale:1999,Chiang:2002,Mustill:2011}. The phenomenon of resonant repulsion has also been investigated \citep{Lithwick:2012,Petrovich:2013,Terquem:2019}.  However, smooth transfer between neighboring first order MMRs is a novel possibility revealed by the low eccentricity resonant bridges. We suggest that it may have interesting applications in the dynamical transport and mixing of small bodies in planetary systems.

\subsection{High eccentricity resonances}\label{ss:high-ecc}

For moderate to high values of $e^*$, we observe the following in Fig.~\ref{f:f3}(a):
\begin{itemize}
\item[(i)]The pericentric branch achieves a maximum width near but slightly below the planet-grazing value, $e_{\mathrm{cross}} =(\frac{p+1}{p})^{\frac{2}{3}}-1$, and then decreases with increasing eccentricity.  Our non-perturbative approach finds that the maximum width is finite, in contrast with some previous studies reporting divergent widths near the planet-grazing eccentricity.
\item[(ii)] The apocentric branch re-emerges at eccentricities exceeding $e_{\mathrm{cross}}$, and grows wider with increasing $e^*$. Both the pericentric and apocentric branches co-exist in the high eccentricity regime.
\item[(iii)]In the case of the MMRs with larger $p$ ($p>1$), there are multiple terminations and re-emergences of pericentric and/or apocentric branches at higher values of $e^*$. %Such features are also reported in our recent studies on Neptune's exterior MMRs in the high eccentricity regime \citep{Malhotra:2018a,Lan:2019,Malhotra:2019b}. 
\end{itemize}

The co-existence of the pericentric and apocentric branches and their terminations/re-emergences at high eccentricities have remained largely unremarked in previous studies.  It is rather remarkable that the widths of resonances at high eccentricities are not much smaller than their maximum width achieved in the lower eccentricity regime. In Section \ref{s:geometry}, we discuss the properties and physical origin of these high-eccentricity features from a geometric point of view.  These features have applications in the high eccentricity populations of minor planets throughout the solar system, and potentially also in exo-planetary systems and multiple star systems.

\cite{Wang:2017} referred to the pericentric and apocentric branches as the ``first resonance zone" and the ``second resonance zone", respectively; later we adopted the terminology of ``pericentric" and ``apocentric" branches, more descriptive of their physical properties. 

\subsection{Dependence on $\mu$}\label{ss:mu}

\begin{figure*}[h]
 \centering
\includegraphics[scale=0.33]{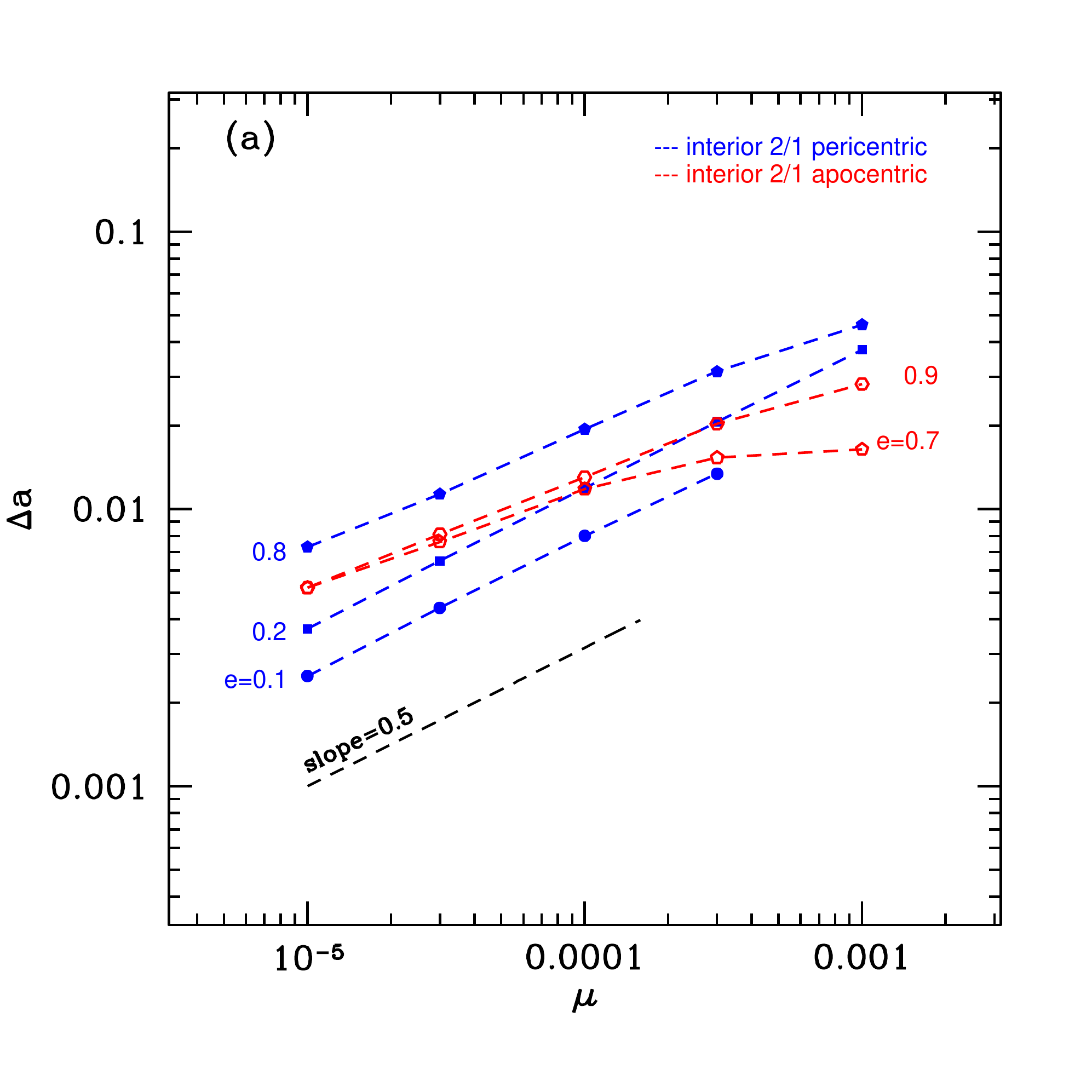} 
 \includegraphics[scale=0.33]{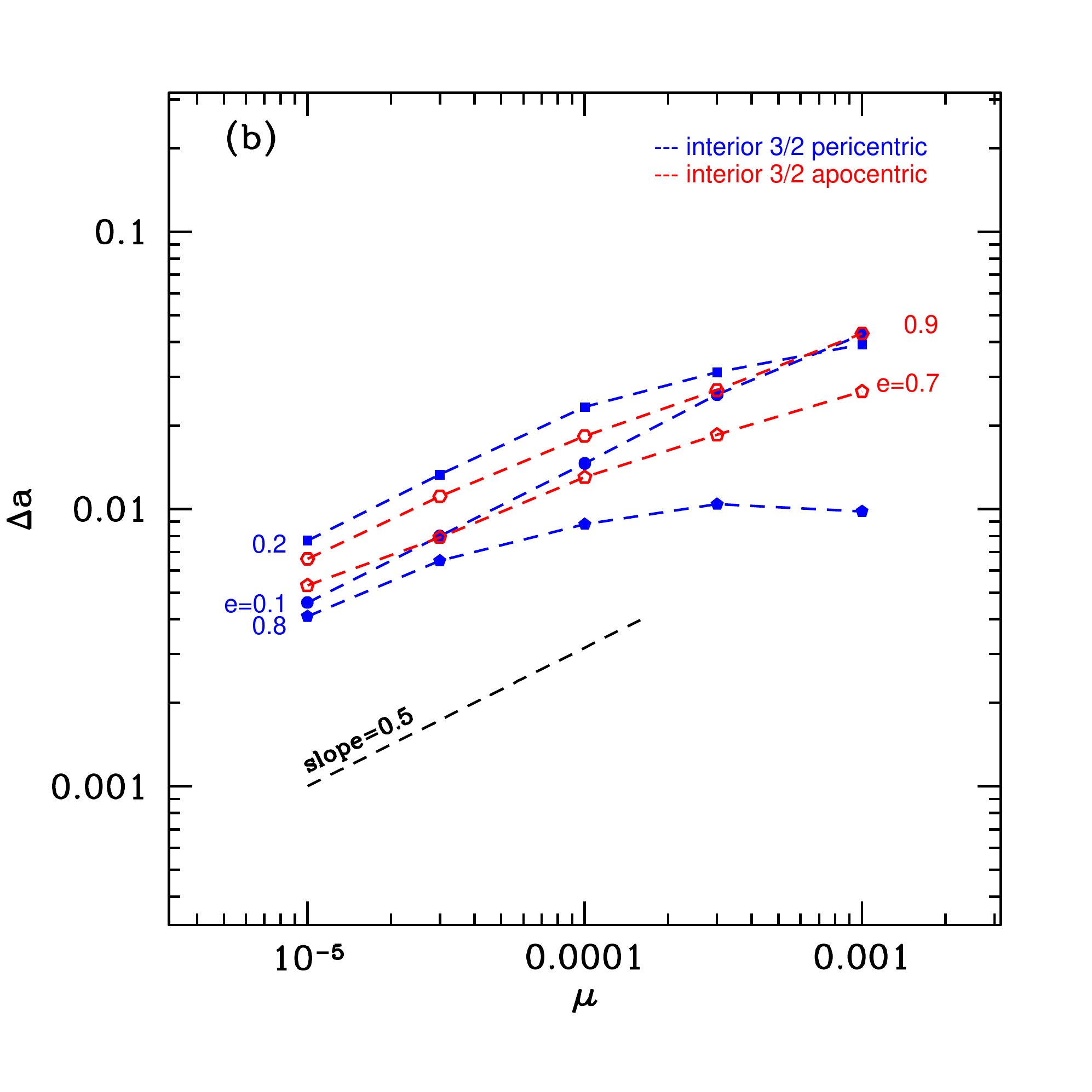}
 \caption{Resonance widths, $\Delta a$, as a function of perturber's mass, $\mu$, (a) for the 2/1 interior MMR, and (b) for the 3/2 interior MMR. The blue curves are for the pericentric branch and the red curves are for the apocentric branch; each curve is for a specific eccentricity of the test particle, as marked on each plot. Based on results reported in \cite{Wang:2017}.
}
 \label{f:f4}
\end{figure*}

A selection of results for the dependence of the resonance widths, $\Delta a$, on the perturber's mass, $\mu$,  is shown in Fig.~\ref{f:f4} for the interior 2/1 and 3/2 MMRs, at a few different eccentricities.  Fitting the measured widths to power laws, $\Delta a \propto \mu^{\beta}$, finds that the pericentric branches at the low eccentricity regime have $\beta\simeq0.5$. In the high eccentricity regime, $\beta\simeq0.4$ for the 2/1 MMR while $\beta\simeq0.33$ for the 3/2 MMR, respectively. The apocentric branches in the high eccentricity regime have $\beta$ in the range 0.3--0.4.  

%%%%%%%%%%%%%%%%%%%%%%%%%%%%%%%%%%%%%%%%%%%%%%%%%%%%
%%%%%%%%%%%%%%%%%%%%%%%%%%%%%%%%%%%%%%%%%%%%%%%%%%%%

In general, guidance from analytical theory leads to the expectation of a sub-linear dependence of resonance widths on the perturber's mass.  For example, from the pendulum model one would expect resonance widths to scale as the square root of the magnitude of the resonant potential. At higher accuracy, the second fundamental model for resonance, SFMR, predicts first order resonance widths to scale as $\sim\mu^{\frac{2}{3}}$ in the low eccentricity regime %, and extrapolation of this model to higher eccentricities predicts the scaling $\sim\mu^{\frac{1}{2}}$ 
\citep{Wisdom:1980,Henrard:1983}. % ,Malhotra:1998}. 
In the high eccentricity regime, there is limited analytical guidance available for the $\mu$--dependence of resonance widths. Extrapolation of the analytical SFMR to higher eccentricities predicts the scaling $\sim\mu^{\frac{1}{2}}$ 
\citep{Malhotra:1998}.  
These perturbation theory estimates of the scalings are somewhat different from the results of our non-perturbative approach. The reasons for the differences are not yet known. 

It is also noteworthy that, in the case of the 2/1 interior MMR, the power laws $\Delta a \propto \mu^{\beta}$ hold well for $\mu$ values up to $0.003$ whereas for the 3/2 interior MMR, significant deviations (toward smaller power law index) are observed for $\mu\gtrsim10^{-4}$. This shallowing of the $\mu$--dependence is likely owed to the crowding of neighboring MMRs, crowding that becomes increasingly significant for larger $\mu$. The surprise is that this crowding appears already for the 3/2 MMR at $\mu\sim10^{-4}$, a value that would not usually be considered large.

\section{Connection of Phase Space Topology to the Geometry of the Resonant Orbit}\label{s:geometry}

\begin{figure*}[hb]
\centering
  \vglue-0.2truein   \begin{tabular}{c} 
         \includegraphics[width=0.45\textwidth, scale=0.7]{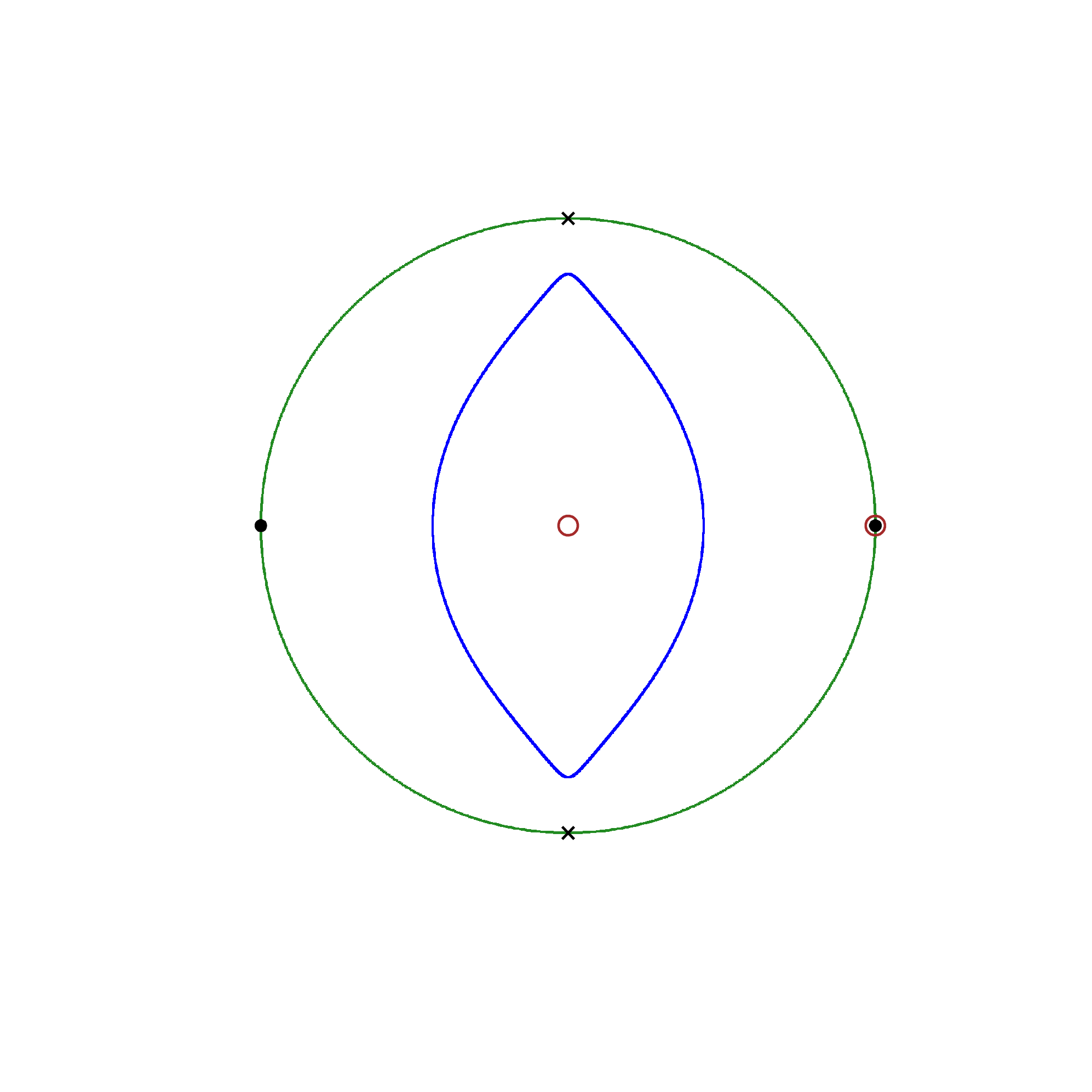}
            \includegraphics[width=0.45\textwidth, scale=0.7]{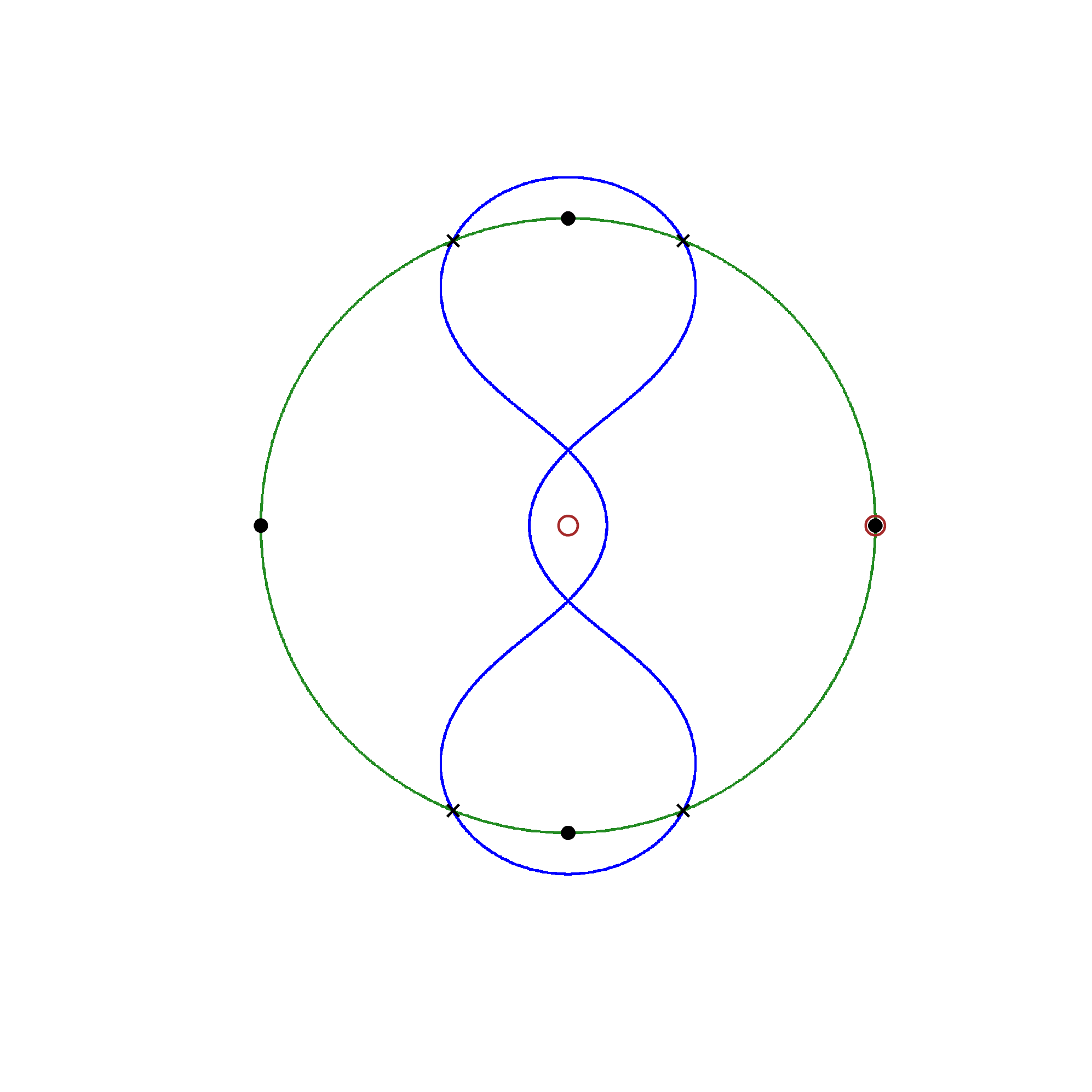}
    \\
           \includegraphics[width=0.45\textwidth]{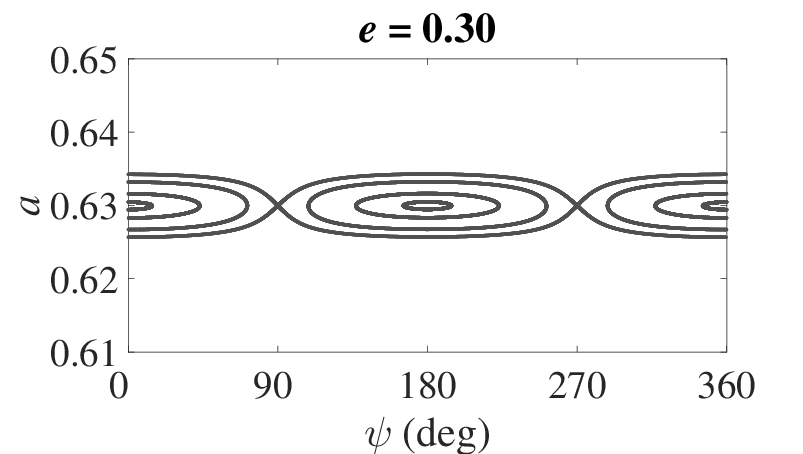} 
        \includegraphics[width=0.45\textwidth]{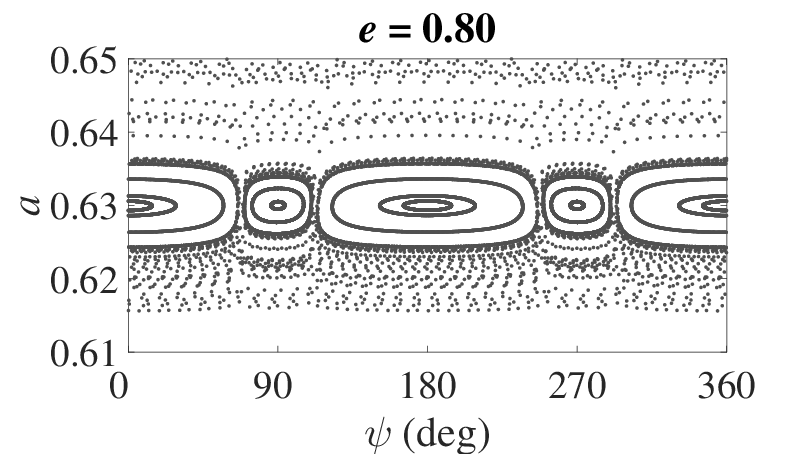}
   \end{tabular} 
\caption{
(top) The geometry of the exact resonant orbit in the rotating frame is traced by the closed curve of two-fold symmetry (shown in blue), for eccentricity 0.3 (left panel) and 0.8 (right panel). The test particle has two perihelion passages in this closed shape, one on the positive abscissa and one on the negative abscissa.  The green circle is of radius $1-\mu$; this would be the path of the planet in the barycentric inertial frame, but in the rotating frame the Sun and the planet's positions are fixed on the abscissa at $-\mu$ and $1-\mu$, respectively, as indicated by the open brown circles. The black dots on the green circle denote the stable orientations of the test particle's perihelion, while `x's  denote its unstable orientations.\\
(bottom) Poincar\'e sections in $(\psi,a)$ near the interior 2/1 MMR, for mass ratio $\mu = 3 \times 10^{-5}$, for eccentricity near $e^*=0.3$ (left panel) and near $e^*=0.8$ (right panel). The chains of resonant islands indicate the stable libration regions surrounding the stable periodic orbit in exact 2/1 resonance with the planet; each periodic orbit appears as a pair of points in the Poincar\'e section. 
Note that the stable and unstable orientations of the resonant orbit (top plots) correspond, respectively, to the centers of the libration islands and to the saddle points in the Poincar\'e sections (bottom plots). The lengths of arcs in-between the stable and unstable orientations correspond to the maximal possible extent in $\psi$ of the libration islands. 
}
\label{f:f5}
\end{figure*}

The geometry of resonant orbits has a fundamental physical connection to the phase space structures and the eccentricity dependence of the resonance widths described in the previous sections. We illustrate this for the interior 2/1 MMR with the help of Figure \ref{f:f5}. In the frame co-rotating with the constant angular velocity of the primaries, and with origin at their center-of-mass, the position of the sun and the planet are fixed at distance $-\mu$ and $1-\mu$ from the origin; these are indicated with the brown open circles in Fig.~\ref{f:f5}(top panels). In this rotating frame, an exact resonant orbit is a periodic orbit: its trace is a closed curve and its shape has $(p+1)$-fold symmetry.  For a specified eccentricity and resonant ratio, there exist multiple periodic orbits; all have the same shape, but are distinguished from each other by their different (but specific) orientations relative to the fixed locations of the sun and planet in the rotating frame.  Each chain of resonant islands in the Poincar\'e section corresponds to a particular orientation of the resonant orbit relative to the (fixed) location of the sun and planet in the rotating frame. 

%%%%%%%%%%%%%%%%%%%%%%%%%%%%%%%%%%%%%%%%%%%%%%%%%%%%
% From Wang:2017
%
%%%%%%%%%%%%%%%%%%%%%%%%%%%%%%%%%%%%%%%%%%%%%%%%%%%%

It is worth stating that non-resonant orbits in proximity to the resonance also trace curves whose shape resembles that of an exact resonant orbit, but their trace does not close; rather, the whole trace gradually rotates all the way around without closing on itself. Librating orbits also trace curves that are not closed, but they oscillate (librate) around the orientation of a stable exact resonant orbit.  The exact resonant orbit is a periodic orbit whereas the librating resonant orbits and the nearby non-resonant orbits are quasi-periodic. A visualization (with animations) of a Pluto-like librating resonant orbit and a non-resonant orbit near Neptune's 2/3 MMR can be found in \cite{Zaveri:2021}.

The closed blue curve in Figure~\ref{f:f5}(top) shows the shape of an eccentric 2/1 interior resonant (periodic) orbit in the rotating frame, for one stable orientation; the left panel shows the case of eccentricity 0.3, the right panel is the case of eccentricity 0.8. The corresponding phase space portraits in the vicinity of these cases are shown in the bottom two panels. We ask the reader to imagine rotating the two-fold symmetric shape of the resonant orbit to different orientations to understand the stability or instability of different orientations that we describe now. This description uses Fig.~\ref{f:f5} as the main aid, but also will refer to the structures of the resonances in the $(a,e)$ plane in Fig.~\ref{f:f3}.

On the green circle of radius $1-\mu$, we denote with a black dot the particle's longitude of perihelion corresponding to each of the stable centers of libration visible in the Poincar\'e section; it is intuitively evident that these orientations are stable because they maximally avoid close encounters with the planet. Indeed, integrated over the trace of the closed blue curve, these orientations minimize the planet's perturbation on the particle. Similarly, on the green circle, we denote with a black `x' the particle's longitude of perihelion corresponding to the geometry of each of the unstable fixed points visible in the Poincar\'e section ((which are saddle points of the potential function). Again, it is intuitively evident that these orientations are unstable because the particle approaches the perturber most closely in these paths. And, integrated over the trace of the closed blue curve, these orientations maximize the planet's perturbation on the particle.

When the eccentricity is not too large, the particle's resonant orbit is completely inside the orbit of the planet and its aphelion is far from the planet. For the case of $\mu=3\times10^{-5}$ and $e=0.3$ illustrated in our example, we see that there is a chain of two libration islands in the Poincar\'e section, and two saddle points at $\psi=90^\circ, 270^\circ$ (Figure \ref{f:f5}, bottom left). A separatrix passing though the saddle points delineates the extent of the resonance libration zone. The two-fold symmetry of the trace of the test particle's resonant orbit in the rotating frame leads directly to the two centers of the stable islands at $\psi=0,180^\circ$ in the corresponding Poincar\'e section. The stable geometry is one in which the test particle's longitude of perihelion is oriented near one of the two locations denoted by the black dots, i.e., with $\psi = 0$, or $\psi = 180^{\circ}$. In this geometry, the planet's perturbation on the particle is minimized.

For larger eccentricity, the aphelion of the test particle gets closer to the orbit of the planet, so the perturbation of the planet on the test particle also becomes larger. Chaotic regions appear in the Poincar\'e section, bounding the stable libration islands. When the eccentricity exceeds $e_{\mathrm{cross}}=(1/2)^{\frac{2}{3}}-1=0.59$, the test particle's orbit is planet-crossing and new stable islands appear in the Poincar\'e section (Figure \ref{f:f5}, bottom right). 
When the aphelion distance of the test particle's orbit exceeds the orbit radius of the planet, the two lobes of the resonant orbit intersect the green circle; the green circle is cut into four arcs. The intersection points, denoted by `x' in the figure, are the collision points. They also represent the new unstable orientations of the test particle's perihelion, and are visible as the new unstable fixed points in the Poincar\'e section. The ranges of longitudes of the two arcs in-between the unstable orientations are where new stable islands appear in the Poincar\'e section. The length of arc of the green circle which is enclosed by each of the two lobes delineates the maximum possible range of $\psi$ for the new resonant islands in the Poincar\'e section, whereas the lengths of arc outside the lobes correspond to the maximum possible range of $\psi$ in the old resonant islands. The chaotic regions in the Poincar\'e section are the perihelion orientations near the `x' points; in these orientations, the particle would have close approaches to the planet. As the eccentricity becomes larger, the length of arc enclosed by each lobe also becomes larger, whereas the length of arc outside the lobes shrinks. So the new stable islands grow and expand at the expense of the range of the old stable islands (Fig.~\ref{f:f3}(a)). In other words, the apocentric islands grow at the expense of the pericentric islands.

As the eccentricity approaches $0.90$, the two lobes intersect each other, and their intersection points get close to the green circle. At this high eccentricity, the trace of the resonant orbit in the rotating frame again defines only two arcs, but these two are the ones enclosed by the lobes.  This means that the original two islands have disappeared and only the new stable islands exist, marking the termination of the pericentric branch in Fig~\ref{f:f3}. At even higher eccentricity, when the self-intersections of the lobes occur outside the green circle, the green circle is again cut into four arcs, so the ``old'' stable islands centered at $\psi=0,180^\circ$ reappear, albeit with smaller sizes. This marks the reappearance of the pericentric branch at high eccentricity in Fig.~\ref{f:f3}(a).

Why are the widths of the different islands in the pericentric branch (respectively, apocentric branch) not all the same (Fig.~\ref{f:f3})? The answer lies in the fact that the planet's perturbation on the particle is not invariant under the transformation $x \rightarrow -x$, but is only invariant under the transformation $y \rightarrow -y$. (Recall that $x,y$ are the coordinates in the rotating frame.) Although the closed shape of the resonant orbit (periodic orbit) nominally appears to be symmetric about the $y$-axis, it is easy to see that this is not an exact symmetry. For example, in the cases of the 2/1 resonant orbits illustrated in Fig.~\ref{f:f5}, the perihelion passage that occurs at $\psi=0$ has smaller distance to the planet than the perihelion passage that occurs at $\psi=180^\circ$. This difference translates into slightly different velocities of the particle at alternate perihelion passages, hence slightly different orbital elements $a^*,e^*$ at the center of the corresponding two pericentric islands visible in the Poincar\'e sections, as well as different values of $a_{\mathrm{min}}$ and $a_{\mathrm{max}}$ at the extrema of the libration zones of each resonant island. This is also the reason for the deviation of the centers of the apocentric islands from exactly $90^\circ$ and $270^\circ$, as noted in Section~\ref{s:new-Poincare}. 

There is an extensive body of literature in periodic orbit theory on families of periodic orbits in the restricted three body problem. From the perspective of periodic orbit theory, the phenomenon of the doubling of the stable islands at some of the transitions in the phase space structure is described as arising from the bifurcation of a periodic orbit from a collision orbit \citep[e.g.,][]{Hadjidemetriou:2000, Voyatzis:2005}. 
Here we have given a physically-intuitive description from a geometric point of view, by reference to the shape and orientation of the resonant orbits in the rotating frame. 

We note that \cite{Wang:2017} described the physical connection between the phase space structures and the shape and orientation of the resonant orbit in the rotating frame from a different point of view: by considering different possible locations of the planet on the green circle, rather than fixing its location on the positive $x$-axis. That explanation as well as the one given here are both correct and indeed equivalent to each other, but one or the other is more useful depending upon the application. For example, allowing the planet's location on the green circle to `float' is of value in predicting the location (within its orbit) of a hypothetical unseen Planet Nine in the distant solar system \citep{Malhotra:2016a}.

\section{Future Directions}

We end this review with a non-exhaustive list of potential future directions for research on this topic.

\begin{enumerate}

%\item While our investigations have resolved some puzzles about the divergence of resonance widths in the previous literature, we have not achieved complete clarity on why our approach reveals the two branches of first order MMRs in the low eccentricity regime whereas the previous approaches did not.  Achieving such clarity awaits a future investigation. 

\item
The connection between the geometry of a resonant orbit and the phase space structure (as revealed in the Poincar\'e sections) is most clear for the regime of moderate-to-high eccentricities; it enables an understanding of resonance dynamics from a geometric and physical point of view. However, in the low eccentricity regime, this geometric point of view and corresponding physical understanding is not so clear. Particularly vexing is the behavior of the apocentric branch of first order interior MMRs (and pericentric branch, in the case of exterior MMRs).
A physical explanation for the low-eccentricity phase space topology awaits future work.

\item
There is much room to investigate further the low eccentricity resonant bridges and their applications to natural systems. These bridges appear continuously over a large range of semi major axis in-between adjacent first order MMRs, but we can imagine that higher order MMRs in-between may play a role as ``off-ramps'' from these bridges. This requires investigation with higher numerical resolution in the Jacobi constant than in the studies published thus far. 

\item[] It is also interesting to examine the relationship between the low eccentricity resonant bridges and the near-resonant and secular approximations. The concepts of ``forced" and ``free eccentricity" in proximity to a mean motion resonance and in secular perturbation theory are usually a good approximation for the regions in-between low order MMRs \citep[e.g.,][]{Murray:1999SSD}. A quantitative comparison between the non-perturbative results and the analytical estimates of forced and free eccentricity awaits future work. 

\item
The locations of the centers, ($a^*,e^*$), of the pericentric and apocentric resonance branches computed in our non-perturbative analysis with Poincar\'e sections have some notable differences compared to those of the corresponding periodic orbits computed by \cite{Antoniadou:2021}. This is particularly visible in the locations of high-eccentricity apocentric branches. A study to reconcile these differences would be potentially useful to clarify the quantitative details of the phase space structure and possibly also to improve numerical algorithms.

\item
For first order interior resonances, the scaling of resonance widths with the perturber's mass found with our non-perturbative approach has some fundamental discrepancies with the analytical guidance from the second fundamental model for resonance (Sec.~\ref{ss:mu}). It would be of value to understand the reasons for these discrepancies. It would also be of value to investigate these scalings for other MMRs, including exterior MMRs and higher order MMRs.

\item
The concept of ``overlapping resonances" as the origin of deterministic chaos has been well accepted for the past few decades. This has provided fairly good quantitative results in applications to the solar system, such as the Kirkwood gaps in the asteroid belt. Some of the new results reported here highlight the role of a distinctly different mechanism, namely bifurcations of resonance zones (equivalently, bifurcations of periodic orbits) near the planet-crossing eccentricity, in generating chaos and reducing or even eliminating stable libration zones. Investigation of the  complementary roles of bifurcations and overlapping resonances could significantly advance our understanding of the origin of dynamical chaos near orbital resonances. 

\item[]
Furthermore, investigation of the transport pathways within the chaotic zones near MMRs has potentially very interesting and useful applications to the migration and mixing of small bodies in planetary systems. 

\item
We have seen that for some ranges of parameters, first order resonances have fuzzy boundaries. Dynamical systems theory explains such fuzzy boundaries as homoclinic tangles arising from perturbations of a separatrix that exists in the one-degree-of-freedom idealization of the system. In the planetary dynamics literature of the past few decades it is usually attributed to overlapping higher order resonances that accumulate near the separatrix. Is there a physical explanation, even an approximate one, as an alternative to the hand-wavy explanation embodied in ``overlapping resonances"? For example, it is conceivable that the stable libration zone boundaries are related to the closest approach distance (and relative velocity at close approach) to the perturber. The scaling of the stable resonance widths with the perturber mass would suggest this to be the case. A nice start in this direction has been made by \cite{Pousse:2021} who have obtained estimates of the limits of the averaged solutions for mean motion resonances related to the close approach distance to the perturber; this approach merits further investigation.
 
\item
Our specific approach to Poincar\'e sections of the PCRTBP has the potential to be extendable to more realistic models for the dynamics of mean motion resonances. The stroboscopic record of state vectors at successive perihelion passages can enable analysis of the dynamics and properties of the phase space neighborhood generated in the non-restricted three body problem, or in non-coplanar models or by multiple planetary perturbers. In these more complex and more-than-two-degrees-of-freedom models, it is still possible to capture the remnants or ``shadows" of the resonant structures with this stroboscopic record in a useful way. By employing so-called {\it pseudo}-Poincar\'e sections inspired by the approach reviewed here, a new study \citep{Volk:2021} reports such an attempt in an investigation of Neptune's exterior MMRs in a three-dimensional spatial setting with a realistic numerical model including the perturbations of all four giant planets, Jupiter, Saturn, Uranus and Neptune.  

\item
Several recent analyses of planetary mean motion resonances in the context of exo-planetary systems have been based on the conventional  critical resonant angle \citep[e.g.,][]{Deck:2013,Ramos:2015,Hadden:2018b,Hadden:2019,Petit:2021}. These could potentially also benefit from validation with the lens of the sub-multiple of the critical resonant angle used in our work, and/or employing physically-motivated stroboscopic records to generate {\it pseudo}-Poincar\'e sections for comparison with the semi-analytic approaches. 

\item
The generation of many Poincar\'e sections and the task of measuring resonance centers and widths by visual examination is labor intensive and fatiguing for a researcher. It would be useful to develop methods to carry out these tasks with machine-learning algorithms. 

\end{enumerate}

\bigskip

\noindent{\it Acknowledgements}

Figures 1--3 and Figure 5 in this paper are adapted from \cite{Malhotra:2020}; Figure 4 is based on results reported in \cite{Wang:2017}. This work was partially supported by research funding from NSF (grant AST-1824869), NASA (grant 80NSSC18K0397), and the Marshall Foundation of Tucson, AZ. 

\bibliographystyle{aasjournal}

\end{document}